\documentclass[12pt]{article}
\usepackage{graphicx}
\usepackage{url}
\usepackage{cite}
\usepackage[margin=1.25 in]{geometry}
\usepackage[colorlinks = true, linkcolor = blue, urlcolor = blue,
      citecolor = blue, anchorcolor = blue]{hyperref}

\newcommand\pubnumber{SLAC-PUB-17701}
\newcommand\pubdate{September 2022}

\def\SLAC{SLAC,
    Stanford University, Menlo Park, California 94025 USA}
\def\doeack{\footnote{Work supported by the US Department of Energy,
                     contract DE--AC02--76SF00515.}}

\def\Title#1{\begin{center} {\Large #1 } \end{center}}
\def\Author#1{\begin{center}{ \sc #1} \end{center}}

\def\submit#1{\begin{center}Submitted to {\sl #1} \end{center}}
\newcommand\pubblock{\rightline{\begin{tabular}{l} \pubnumber\\
         \pubdate \end{tabular}}}
\newenvironment{Abstract}{\begin{quotation} \begin{center}
                       ABSTRACT
     \end{center}\bigskip  }{\end{quotation}}

\def\submit#1{\begin{center} {\sl Submitted to #1} \end{center}}
\def\Acknowledgements{\bigskip  \bigskip \begin{center} \begin{large}
             \bf ACKNOWLEDGEMENTS \end{large}\end{center}}

\def\beq{\begin{equation}}
\def\eeq#1{\label{#1}\end{equation}}
\def\eeqn{\end{equation}}

\newenvironment{Eqnarray}%
   {\arraycolsep 0.14em\begin{eqnarray}}{\end{eqnarray}}
\def\beqa{\begin{Eqnarray}}
\def\eeqa#1{\label{#1}\end{Eqnarray}}
\def\eeqan{\end{Eqnarray}}
\def\CR{\nonumber \\ }

\def\leqn#1{(\ref{#1})}

\let\bar=\overbar

\def\VEV#1{\left\langle{ #1} \right\rangle}

\def\lsim{\mathrel{\raise.3ex\hbox{$<$\kern-.75em\lower1ex\hbox{$\sim$}}}}
\def\gsim{\mathrel{\raise.3ex\hbox{$>$\kern-.75em\lower1ex\hbox{$\sim$}}}}

\def\L{{\cal L}}

\def\L{{\cal L}}

\def\half{\frac{1}{2}}

\def\del{\partial}
\def\Dslash{\not{\hbox{\kern-4pt $D$}}}
\def\dslash{\not{\hbox{\kern-2pt $\del$}}}

\def\Dlr{\mathrel{\raise1.5ex\hbox{$\leftrightarrow$\kern-1em\lower1.5ex\hbox{$D$}}}}

\def\ee{e^+e^-}

\def\msb{{\bar{\scriptsize M \kern -1pt S}}}

\def\drb{{\bar{\scriptsize D \kern -1pt R}}}

\def\eps{\epsilon}

\makeatletter
\def\section{\@startsection{section}{0}{\z@}{5.5ex plus .5ex minus
 1.5ex}{2.3ex plus .2ex}{\large\bf}}
\def\subsection{\@startsection{subsection}{1}{\z@}{3.5ex plus .5ex minus
 1.5ex}{1.3ex plus .2ex}{\normalsize\bf}}
\def\subsubsection{\@startsection{subsubsection}{2}{\z@}{-3.5ex plus
-1ex minus  -.2ex}{2.3ex plus .2ex}{\normalsize\sl}}

\renewcommand{\@makecaption}[2]{%
   \vskip 10pt
   \setbox\@tempboxa\hbox{\small #1: #2}
   \ifdim \wd\@tempboxa >\hsize     
       \small #1: #2\par          
     \else                        
       \hbox to\hsize{\hfil\box\@tempboxa\hfil}
   \fi}

\makeatother

\begin{document}
\begin{titlepage}
\pubblock

\vfill
\Title{Model-Agnostic Exploration of the Mass Reach \\ of Precision Higgs
  Boson Coupling Measurements}
\vfill
\Author{ Michael E. Peskin\doeack}
 \medskip
\begin{center} 

  \SLAC 
\end{center}
\vfill
\begin{Abstract}
To understand the possibility for precision Higgs boson coupling
measurements to access effects of very heavy new particles, I present
five scenarios in which significant deviations in Higgs boson
couplings are produced by new particles with multi-TeV masses.   These
scenarios indicate that such precision measurements provide
opportunities to reach well
beyond the capabilities of direct particle searches at the HL-LHC.
\end{Abstract} 
\vfill
\submit{the Proceedings of US Community Study\\ 
on the Future of Particle Physics (Snowmass 2021)}

\vfill
\end{titlepage}

\hbox to \hsize{\null}


\tableofcontents

\def\thefootnote{\fnsymbol{footnote}}
\newpage
\setcounter{page}{1}

\setcounter{footnote}{0}

\section{Introduction}

One of the goals of a program of precision measurements on the Higgs
boson is to prove  that the Higgs boson is affected by new physics.
Ideally, these precision measurements should indicate the existence of
new, undiscovered particles.  These precision measurements would be
especially powerful if they can be sensitive to new particles that are well beyond the
search reach of the High Luminosity Large Hadron Collider (HL-LHC). 

Unfortunately, it is not so straightforward to test this claim. 
   Models of Beyond-Standard-Model (BSM) physics
typically contain many parameters that can be adjusted.  It is always
possible to cherry-pick points in parameter space that give very
strong 
sensitivity to particles of high mass.  Examples are cited
in~\cite{Barklow:2017suo,ILCInternationalDevelopmentTeam:2022izu}.
  But it may not be clear to what extent these points are
generic rather than exceptional.

One solution is 
to restrict oneself to  the same subsets of BSM
model space that are examined at the LHC.  However, this hides an
important strength of precision Higgs measurements.  In complex BSM models,
the studies done by the
LHC experiments often rely on particular scenarios that are
favorable for direct particle discovery.
But a feature of precision measurements is that they often give access
to a different class of models than those for which the 
the lowest-mass BSM particles are light enough to be
discovered at the LHC.  This complementarity has been noticed in
supersymmetry models, in scans of the PMSSM parameter
set such as ~\cite{Cahill-Rowley:2014wba}.  However, such multi-parameter
scans are reliant on public codes that may not be tested over the whole
parameter region of interest.  For example, this particular study
uses a version of the code {\tt HDECAY}~\cite{Djouadi:1997yw}
 that is not accurate for
very heavy squark masses.  In general, it is very difficult to check
that public codes for complex models 
written with low-mass BSM particles in mind are
valid uniformly over the whole parameter space.

In this note, I approach this question in a different way. I
identify specfic scenarios in which integrating out individual  heavy
particles or heavy particle sectors
leads to potentially large corrections to the Higgs boson
couplings.  I base my analysis on Standard Model Effective 
Field Theory (SMEFT).   I  present the Wilson coefficients generated in 
SMEFT  for the leading corrections to the 
Standard Model Higgs predictions.   These expressions
isolate a small number of parameters
that are important to each particular scenario.   For each case,
this leads  to a small parameter space that can be explored
comprehensively.   These  scenarios then indicate opportunities
for the discovery of new physics  up to high mass scales by Higgs boson precision
measurements.  The scenarios I discuss are simple  representations of
the physics of more complicated BSM models. In fact, they are
deliberately oversimplified to make the specific physics mechanism
clear.  In each case, this points to a path to find these mechanism
working in more complete but more complicated models.

After some general orientation, I will explore the following
scenarios:  (a) the Strongly Coupled Light Higgs (SILH), (b)
two-Higgs-doublet (THDM) models, treated at the tree level; (c) Higgs field
mixing with a scalar singlet boson field ; (d) integration out 
of a heavy vectorlike quark; (e) an effect of the stop-higgsino system
that appears in the Minimal Supersymmetric Standard Model.

For each case, I will give a plot of the reach in the mass of the new
particles in TeV.   Note that I quote 3~$\sigma$ and 5-$\sigma$
discovery reach, not exclusion limits. The projected discovery reach
of the HL-LHC improves with time, but it is generally expected that
the LHC cannot discover new pair-produced new particles with masses as
high as 2~TeV, except in cases with very high cross sections and
striking signatures, for example, a gluino pair decaying to 4 heavy
flavor jets plus missing energy.  For example, in
\cite{CidVidal:2018eel}, the discovery reach at the HL-LHC for vectorlike quarks is
estimated not to exceed 1.5~TeV and the reach for top squarks is
estimated to be 1.4~TeV.  On the other hand, we  will see that precision Higgs boson 
coupling measurements can be sensitive to these particles at masses
well above 2~TeV.

\section{General orientation}

  In SMEFT, the effective Lagrangian describing the Higgs boson
  and its decays is taken to be
\beq
   \L = \L_{SM} + \sum_{J}   {c_J\over v^{D_J-4}} {\cal O}_J 
\eeqn
where $\L_{SM}$ is the Lagrangian of the Standard Model (SM),  the
${\cal O}_J$ are higher-dimension operators respecting the symmetries
of the SM, generated by integrating
out heavy fields, and the $c_J$ are  dimensionless numbers, the Wilson
coefficients.  I take $v$, the Higgs field vacuum expectation value,
to be the reference mass scale for this expansion.  
The leading corrections to the Higgs boson
couplings are given by the operators of dimension $D_j = 6$.
The coefficients of these operators are proportional to
$M^{-2}$, where $M$ is the mass of the heavy particles, and so the 
corrections to the corresponding $c_j$ and to the dimensionless Higgs
boson couplings  
are of the order of $v^2/M^2$.

  A naive estimation of the corrections to the Higgs boson couplings
might be obtained by setting $M = 2$~TeV and 
putting the coefficient 1 in front of
the dimensional estimate.  This gives the size of these effects
as
\beqa
\mbox{tree\ level\ effects:} &\quad &  v^2/M^2 \sim  1\%  \CR 
  \mbox{loop\ level\ effects:} &\quad & (g^2/4\pi)  v^2/M^2 \sim
  0.1\%
  \eeqan
 Since the capability of proposed $\ee$ Higgs factories is to measure
  the Higgs boson couplings to precisions of order 1\%, these
  estimate suggest that the mass reach of Higgs precision is small.

  However, in many cases, the coefficient
  in front of the parametric dependence can be a large dimensionless
  number. Those cases give opportunties
 for the precision measurement of Higgs
  couplings to be an especially powerful route to discovery.   In this
  paper, I study a few concrete examples of systems
  that generate such large couplings.

For most of the examples, I will discuss  only the corrections to
Higgs boson couplings that can be studied at the low-energy stages of
$\ee$ Higgs factories, at center of mass energies close to 
250~GeV, 360~GeV, and 500~GeV.  I will
ignore the dimension 4 and 6 operators involving the top quark
field.  For definiteness, the plots below will 
compare the deviations expected in the various scenarios to the  
the projected errrors in Higgs couplings expected for the 
500~GeV  International Linear Collider
(ILC)~\cite{ILCInternationalDevelopmentTeam:2022izu,LCCPhysicsWorkingGroup:2019fvj}.
Details of the SMEFT analysis for ILC projections are given in 
\cite{Barklow:2017suo,Barklow:2017awn}.
The measurement of the Higgs self-coupling, which is possible
at 500~GeV and above, can add to the evidence for  new physics, but
that effect will not be included in the plots.  

 Note that whenever we add dimension-6 operators to the SM Lagrangian,
 we must also modify the SM parameters, which are fit to data.  I do
 this by shifting the basic parameters $v$, $g$, $g'$, $\lambda$ and
 the quark and lepton Yukawa couplings
 so that the predicted values of $\alpha(m_Z)$, $G_F$, $m_Z$, $m_h$,
 and the fermion masses $m_f$
 remain unchanged.

\section{Strongly Interacting Light Higgs}  

 One set of
  models that is known to give large coefficients is that  in which the Higgs
  boson is assumed to be a composite particle bound by new strongly
  interacting forces.  A simple framework for the SMEFT Wilson
  coefficients generated by this hypothesis is the Strongly
  Interacting Light Higgs (SILH) model, suggested in~\cite{Giudice:2007fh}.  
This model is
  characterized by a coupling $g_*$, which is assumed to be strong,
  and a scale $\Lambda$ of  resonances associated with the strong
  interactions.
  I take $H$ to be the effective Higgs
  doublet field in the low-energy Lagrangian.  Then the $c_j$
  for dimension-6 operators 
  naturally contain factors $v^2/\Lambda^2$.  The largest contributions come from
  the operators~\cite{Henning:2014wua}
  \beq
  \Delta \L =   { c_H\over 2v^2}\ (\del_\mu (H^\dagger H))^2 
+ { c_b \over v^2 }\ |H|^2
  \, Q_L\cdot H b_R + h.c. \ ,
  \eeqn
for which
  \beq
  c_H  = c_b =   g_*^2 {v^2\over \Lambda^2}
  \eeqn

\begin{figure}
\begin{center}
\includegraphics[width=0.8\hsize]{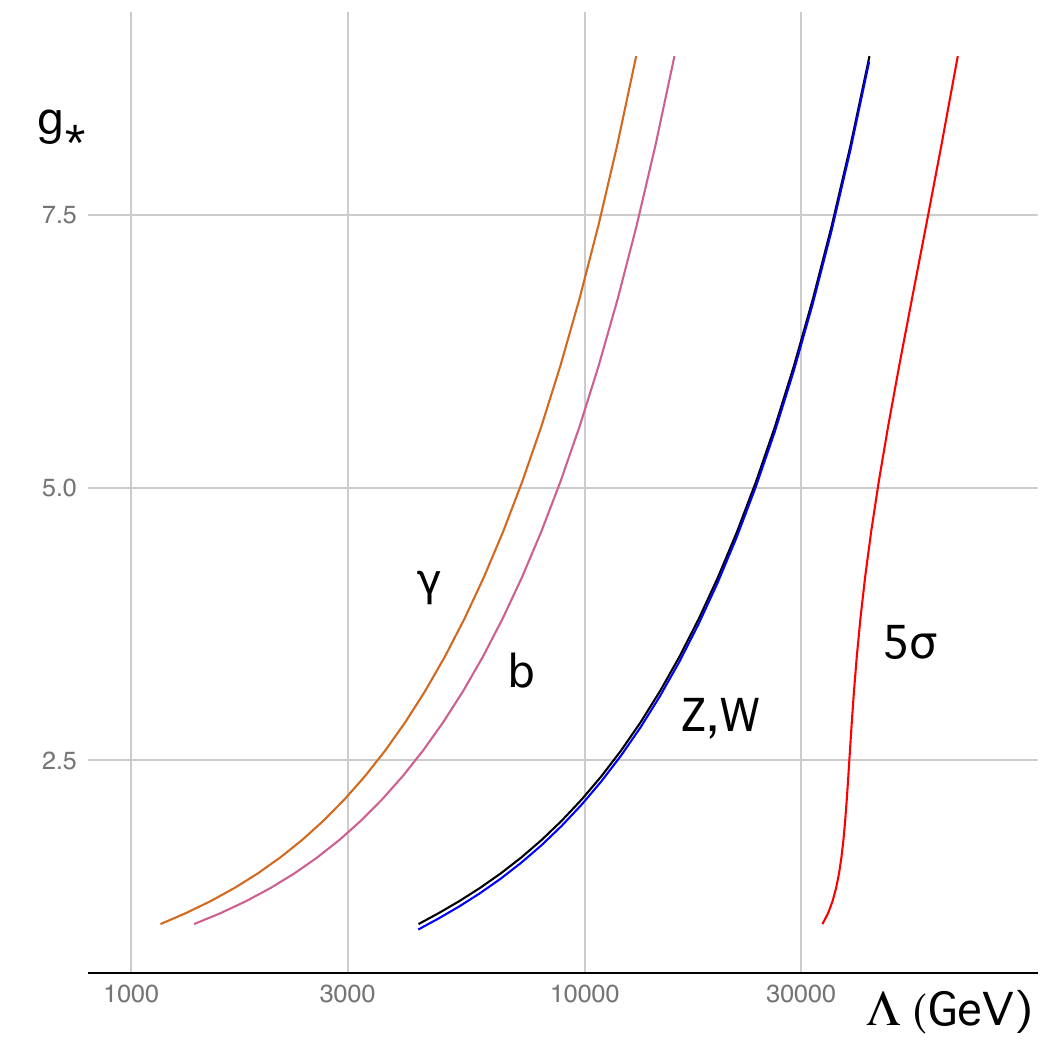}
\end{center}
\caption{Mass reach for the deviation in the Higgs boson couplings
  Higgs compositeness, as described by the SILH  model.   The blue,
  black, magenta, and chocolate curves indicate the 3~$\sigma$
  contours for the deviation in the $HWW$, $HZZ$, $Hgg$, and $H\gamma\gamma$ couplings 
in a SMEFT fit to projected ILC
  data up to 500~GeV.   The curves are plotted in the $(\Lambda,g_*)$
  plane. 
  The
  red curves show the 5~$\sigma$ contour for the full SMEFT fit.}
\label{fig:SILH}
\end{figure}

    In Fig.~\ref{fig:SILH}, I plot the contours in the ($\Lambda, g_*$)
plane along which individual Higgs boson couplings have a
  3~$\sigma$ deviation from the SM expectation, assuming
  measurements with the uncertainties expected for the 500~GeV ILC.
  Other Higgs factory proposals give very similar expectations.  
This figure is very similar to Fig.~8.4 of
\cite{EuropeanStrategyforParticlePhysicsPreparatoryGroup:2019qin},
where the limits for the various Higgs factories are displayed
separately.   The
  combination of these effects and effects of other dimension-6 SMEFT
  operators  gives the contour for a
  5~$\sigma$ deviation from the SM which is also shown in
  the figure. (Here and below, the distance in $\sigma$ between the new
  model and the SM is
  computed as in Sec.~7 of \cite{Barklow:2017suo}.)   The significance
  of the deviations is boosted at small values of $g_*$
  by contributions from deviations in  precision electroweak observables.

  \section{Two Higgs doublet models}  

A set of models that
  gives corrections to the Higgs boson couplings at the tree level is
  built by adding to the SM
  a second Higgs doublet.   The literature on Two
  Higgs Doublet models (THDM) is vast.   Recently, there has been
  much interest in the {\it alignment limit}~\cite{Bernon:2015qea}, in 
which the mixing has very
  small effects on the couplings of the 125~GeV Higgs boson while the
  new bosons remain light enough to observe at the LHC.  Here I will
  consider the {\it decoupling limit}~\cite{Haber:1994mt}, where the new bosons are very
  heavy, perhaps out of reach of the LHC.   For simplicity, I include
  only the leading correction for large masses of the new Higgs
  bosons.  A more complete discussion of this limit has been given in 
\cite{Egana-Ugrinovic:2015vgy,Belusca-Maito:2016dqe}.

  To make the description of the decoupling limit more transparent, I
 begin with a description in terms of a light doublet  field $h$
  and a heavy doublet field $\Phi$.  The most general renormalizable Lagrangian for
  these fields is
  \beqa
  \L &=& |D_\mu h|^2 + \mu^2 |h|^2 - \lambda |h|^4 +  |D_\mu \Phi|^2
  - M^2 |\Phi|^2 \CR
     &  &   \hskip 1.0in  + \eta |h|^2 (h^\dagger \Phi + \Phi^\dagger h)  +
  \cdots \ ,
  \eeqa{simpleTHDM}
  with $\mu^2 \ll M^2$.  I consider the case $M^2 > 0$, so that the 
heavy field does not obtain on its own a large $SU(2)\times
U(1)$-breaking expectation
value. The omitted terms are 
 proportional to $h^2\Phi^2$ and higher powers of $\Phi$.
  For the leading $v^2/M^2$  effect at the tree level, only the terms
  shown explicitly in 
  \leqn{simpleTHDM} are relevant.  

The field $h$ acquires a vacuum
  expectation value $\VEV{h} = (0, v_h)/\sqrt{2}$ where  $v_h$ is
  very close to  the SM expresson $\sqrt{\mu^2/\lambda}$.  The
    coupling $\eta$ generates a  small mixing between $h$ and $\Phi$
    that induces  $\VEV{\Phi} = (0, v_\Phi)/\sqrt{2}$, where
    \beq
    v_\Phi = \eta { v_h^2\over 2 M^2}   \ .
    \eeqn
  One linear combination of $h$ and $\Phi$ has the vacuum expectation value.  This is
  reflected in the Goldstone boson mass eigenstate 
  \beq
   \pi^0 = \pi_h^0 + {\eta v^2/2 \over  M^2} \pi^0_\Phi \ .
    \eeqn
    On the other hand, the mass eigenstate of the light CP-even boson
    is
   \beq
   H^0  = h^0 +  {2\mu^2 + 3\eta v^2/2 \over  M^2} h^0_\Phi \ . 
   \eeqn
  The 125~GeV Higgs boson mass $m_H^2$ equals $2\mu^2$ to this
   order.  Then  the small angle between these vectors,
   \beq
   \gamma = {m_H^2 + \eta v^2\over M^2} \ , 
   \eeq{gammaval}
   is directly observable as a deviation of the Higgs boson couplings
   from the Standard Model expectation $   y_f =  \sqrt{2} m_f/  v $.

In the THDM at the tree level, the deviation in the Higgs
couplings to $W$ and $Z$ is of order $v^4/M^4$.   The $W$ and $Z$
couplings do receive leading contributions in other extended Higgs
sections, for example, those discussed in the next section.

   In a typical THDM, discrete symmetries force one linear combination
   of $h$ and $\Phi$ ($\Phi_1$) to couple to one set of quarks
   and leptons, with the orthogonal linear combination $\Phi_2$
   coupling to a different set of quarks and leptons.  For example, in
   the Type II THDM found in supersymmetric models, $\Phi_1$ couples
   to the down-type  quarks and charged leptons while $\Phi_2$ couples to
   the up-type quarks.  Define
   \beq
   \tan \beta =   \VEV{\Phi_2}/\VEV{\Phi_1}  \ ,
   \eeq
   and distinguish the two fields by restricting to  $\tan\beta > 1$.
   Then the relative corrections to the Higgs couplings for fermions
 that receive their masses from  $\Phi_1$ are
   \beq
   { \Delta y_f\over y_f} = - \gamma \cdot \tan \beta
   \eeqn
   while the relative corrections to the Higgs couplings for fermions
  for fermions
 that receive their masses from  $\Phi_2$ are
   \beq
   { \Delta y_f\over y_f} = \gamma / \tan\beta \ .
   \eeqn  

In supersymmetric models, the value of $\eta $ is typically small.
For example, the contribution to $\eta$ from the $SU(2)$ D-term is 
\beq
   -  {g^2\over 8}\,  \sin2\beta \, \cos 2\beta \to    0.1 /\tan \beta 
\eeqn
in the large $\tan \beta $ limit.  In more general THDMs, $\eta$ can
be larger, up to $\eta \sim 1$.    

\begin{figure}
\begin{center}
\includegraphics[width=0.8\hsize]{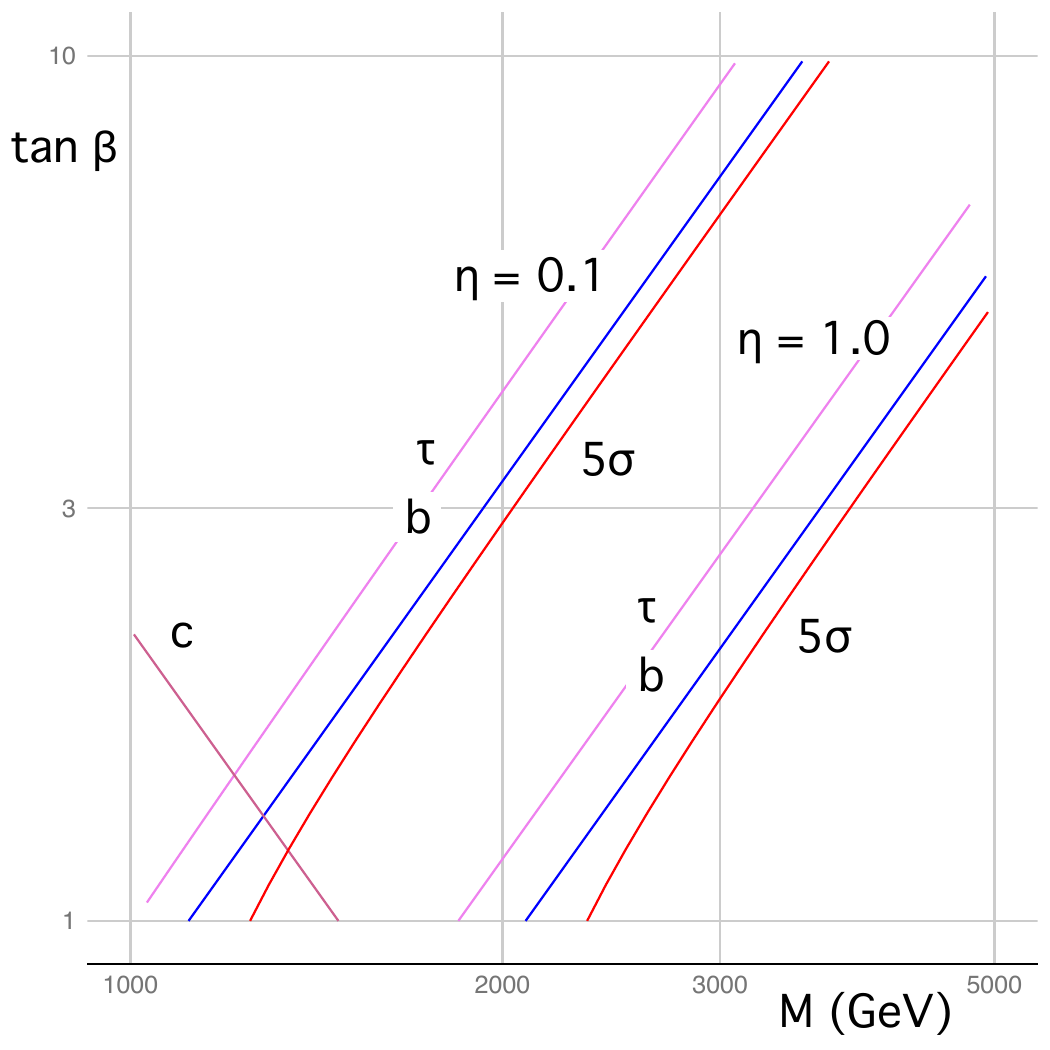}
\end{center}
\caption{Mass reach for the deviation in the Higgs boson couplings due to the
mixing of the Higgs boson with a second Higgs doublet.
  The blu and magenta curves indicate the 3~$\sigma$
  contours for the deviation in the $Hbb$, $H\tau\tau$ couplings 
in a SMEFT fit to projected ILC
  data up to 500~GeV.   The curves are plotted in the $(M, \tan\beta)$
  plane for $\eta  = 0.1$ (left curves) and $\eta = 1.0$ (right
  curves).  
  The
  red curves show the 5~$\sigma$ contour for the full SMEFT fit.}
\label{fig:THDMII}
\end{figure}

\begin{figure}
\begin{center}
\includegraphics[width=0.8\hsize]{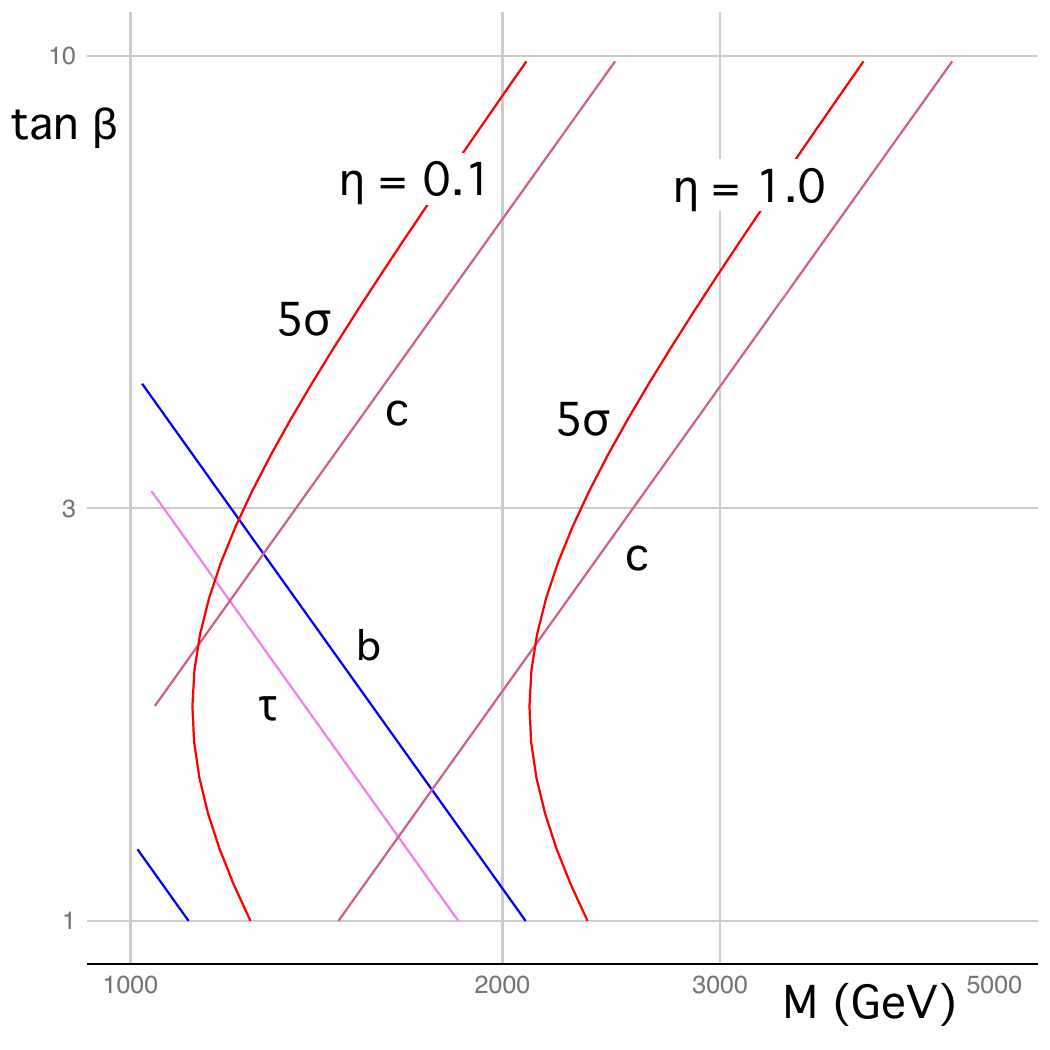}
\end{center}
\caption{Mass reach for the deviation in the Higgs boson couplings due to the
mixing of the Higgs boson with a second Higgs doublet.
  The blue, magenta, and pink curves indicate the 3~$\sigma$
  contours for the deviation in the $Hbb$, $H\tau\tau$, and $Hcc$ couplings 
in a SMEFT fit to projected ILC
  data up to 500~GeV.   The curves are plotted in the $(M, \tan\beta)$
  plane for $\eta  = 0.1$ (left curves) and $\eta = 1.0$ (right
  curves).  
  The
  red curves show the 5~$\sigma$ contour for the full SMEFT fit.}
\label{fig:THDMc}
\end{figure}

    In Figs.~\ref{fig:THDMII} and \ref{fig:THDMc}, I plot the contours in the ($M, \tan\beta$)
 plane along which individual Higgs boson couplings have a
  3~$\sigma$ deviation from the Standard Model expectation, assuming
  measurements with the uncertainties expected for the 500~GeV ILC.
   The
  combination of these effects and other effects of dimension-6
  operators
 gives the contour for a
  5~$\sigma$ deviation from the Standard Model which is also shown in
  the figure.
 Figure~\ref{fig:THDMII} shows the reach for the
  standard Type II model in which the Yukawa couplings of the $b$ and
  $\tau$ are enhanced by $\tan\beta$.  In complete supersymmetric
  models, the expected mass reach at the HL-LHC for the heavy Higgs boson is below
  1~TeV for low values of $\tan\beta$~\cite{CMS:2019qzn}, leading to  a parameter
  region where the precision Higgs reach extends beyond this.
 That region is broader for models with larger values of
  $\eta$.  Figure~\ref{fig:THDMc} shows
  the corresponding situation for a flavorful THDM in which the Higgs coupling to charm
 is enhanced by $\tan\beta$.

 \section{Higgs field mixing with a singlet scalar field}

 It is also possible to add to the Standard Model a heavy singlet Higgs
 field coupling to the 125~GeV Higgs boson through
 \beqa
  \L &=& |D_\mu h|^2 + \mu^2 |h|^2 - \lambda |h|^4 +  |D_\mu \Phi|^2
  - M^2 |\Phi|^2  - {\mu_\Phi\over 3!} \Phi^3 - {\lambda_\Phi\over 4!} \Phi^4 \CR
     & & \hskip 0.2in -  \ A |h|^2 \Phi - k|h|^2 \Phi^2 + \cdots  \ .
  \eeqa{simplesinglet}
Note that $A$ has the dimensions of GeV; I will treat $A/M$ as being order 1.
Such a field can substantially modify the Higgs potential, leading to the
possibility of a first-order electroweak phase transition, while
maintaining smaller effects on the Higgs boson couplings to fermions
and vector bosons.   In  \leqn{simplesinglet}, the cubic $A$ term
gives the dominant effect.   This is a mixing of the singlet Higgs
field into the 125~GeV Higgs boson  with mixing angle
\beq
          \gamma =  {A v\over M^2} .
\eeqn
For $A\sim M$, $v \ll M$, this effect appears as a uniform decrease in
Higgs boson couplings by the factor
\beq
           \cos \gamma  \sim  1 - \half \gamma^2 \sim 1 -\half  {A^2v^2\over
             M^4}  \ . 
\eeqn
In 
models with a $Z_2$ symmetry that forbids this term, substantial  reach for
effects of the singlet field can still be obtained through
 loop effects; see ~\cite{Banta:2021dek}.

The largest effects of the Higgs-singlet mixing are seen in the
self-coupling and in $c_H$~\cite{Henning:2014wua},
\beq 
c_H =  {A^2 v^2 \over M^4}   \qquad    c_6 = - {k A^2 v^2\over 2 \lambda
  M^4} - {\mu_\Phi A^3 v^2 \over  3! \,\lambda M^6 }
\eeqn
Again, note that both expressions are of order $1/M^2$. 
It is important that  $c_H$ modifies all Higgs boson couplings equally, so its
 effects can only be seen by measurements sensitive to the absolute
 normalization of Higgs couplings.

\begin{figure}
\begin{center}
\includegraphics[width=0.8\hsize]{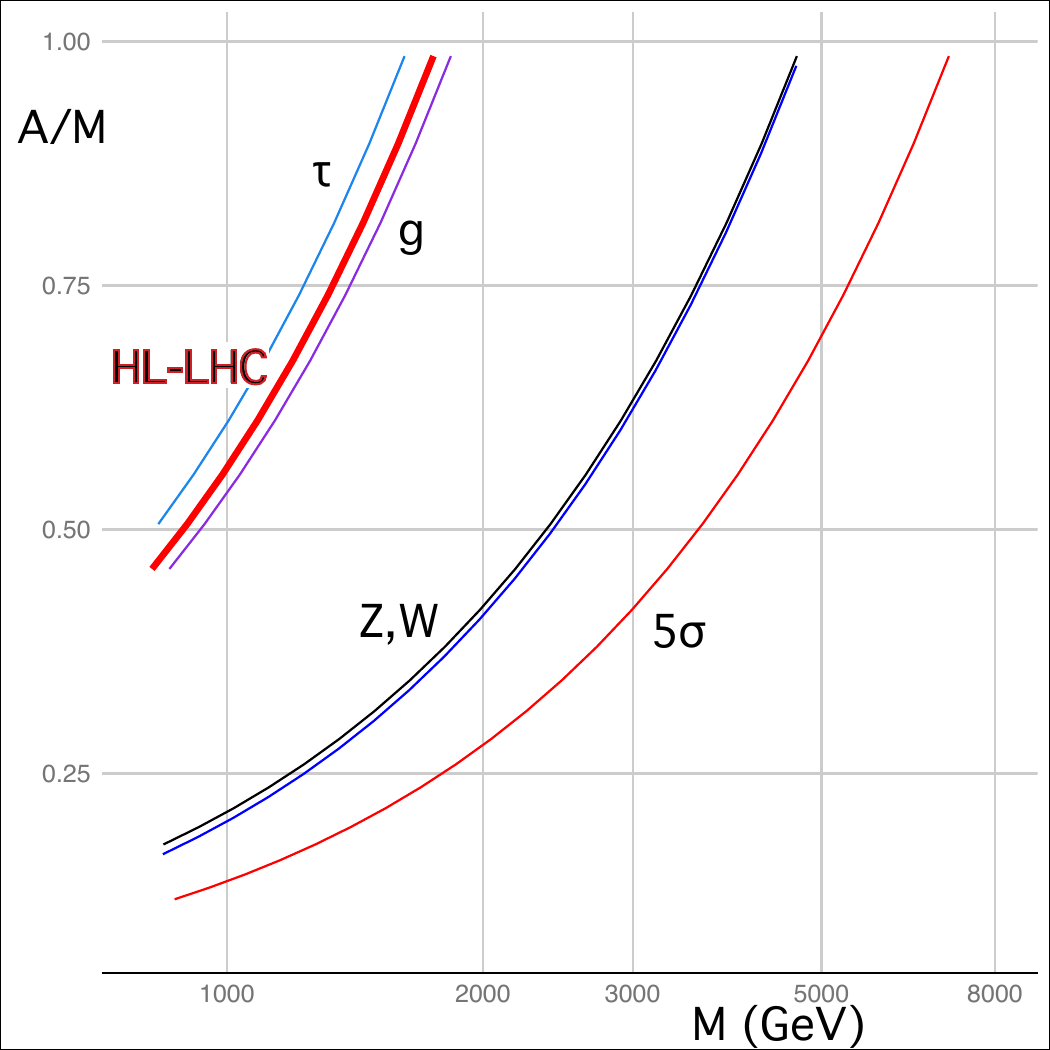}
\end{center}
\caption{Mass reach for the deviation in the Higgs boson couplings due
  Higgs mixing with a heavy scalar singlet.   The blue, black, 
  violet, and light blue curves indicate, respectively,  the 3~$\sigma$
  contours for the deviation in the $HWW$, $HZZ$, $Hgg$,
  and $H\tau\tau$  couplings 
in a SMEFT fit to projected ILC
  data up to 500~GeV.      The curves are plotted in the $(M,A/M)$
  plane.
  The
  red curve shows the 5~$\sigma$ contour for the full SMEFT fit. The
  heavy red line shows the 5~$\sigma$ contour that would be obtained
  from an HL-LHC 1~sigma uncertainty of 3\% on the overall signal strength for Higgs
  processes.}
\label{fig:singlet}
\end{figure}

    In Fig.~\ref{fig:singlet}, I plot the contours in the ($M, A/M$)
plane along which individual Higgs boson couplings have a
  3~$\sigma$ deviation from the Standard Model expectation, assuming
  measurements with the uncertainties expected for the 500~GeV ILC.
 The
  combination of these effects and others gives the contour for a
  5~$\sigma$ deviation from the Standard Model which is also shown in
  the figure.  For comparison, the heavy red line gives the 5~$\sigma$
  contour corresponding to a 3\% uncertainty in the overall signal
  strength for Higgs reactions that might optimistically be measured
  at the HL-LHC.  This corresponds  to a 1.5\% uncertainty
  in the scale of Higgs boson couplings.  Of course, the LHC
  determination has more model-dependence than that obtained from 
$\ee$ Higgs factories. 

It should be noted that there is search reach at the HL-LHC for
  singlet Higgs bosons up to a mass of 2.5~TeV, but only in models with 
very large and visible Higgs-singlet mixing; see, for example, Fig.~54
of  \cite{CLIC:2018fvx}.   The relevant parameter region is above the
top of the plot in Fig.~\ref{fig:singlet}. 

The effect on the Higgs self-coupling is larger than
  that shown here by a factor
\beq 
         1 / \lambda \sim  {\cal O}(10) \ .
\eeqn
 That is, few-percent-level changes in the $HWW$ and
$HZZ$ couplings can potentially  signal order-1 changes in the Higgs
self-coupling, though there might be compensating factors.   The
relation between the  deviations in Higgs couplings to $W$ and $Z$ and
those in the self-coupling are studied in explicit 
models in \cite{Huang:2016cjm,DiVita:2017eyz,Durieux:2022hbu}. 
Note that I do not include the effect of a deviation in the Higgs
self-coupling in the discovery significance plotted in Fig.~\ref{fig:singlet}.

\section{Models with vectorlike quarks}

Integration out of heavy new particles leads to loop corrections to
the Higgs boson couplings.   These are naturally of the order of 
\beq
                          { g^2\over 4\pi}\  {v^2\over M^2}
\eeqn
and often are at the parts-per-mil level.  However, there are
exceptions.  These occur when the loops acquire large factors from
coupling constants, combinatoric factors, factors of $\tan\beta$, or large mass ratios.

The simplest context in which this occurs is the integration out of a
doublet of heavy vectorlike quarks.  I consider the multiplet (Q, U,
D), where $Q$ is an $SU(2)$ doublet and $U$ and $D$ are $SU(2)$
singlets with hypercharges $2/3$ and $-1/3$.  The Lagrangian is 
\beqa
\L &=&  \bar Q (\Dslash - M) Q  + \bar U (\Dslash - M) U +  \bar D
(\Dslash - M) D  \CR
  &=&  \hskip 0.1in  + ( y_D  \bar Q \cdot H\, D   + y_U \bar Q \cdot \widetilde
  H\,  U + h.c.)
\eeqan
where $\widetilde H_a = \eps_{ab} H^*_b$. For simplicity, I have taken
the masses of the heavy quarks to be equal.  I ignore the couplings of
this doublet to light quarks, which must be treated model by model.

 The largest contributions in this case come from
  the operators~\cite{Angelescu:2020yzf}
  \beq
  \Delta \L =   { c_H\over 2v^2} \ (\del_\mu (H^\dagger H))^2  + {g_s
    c_{GG} \over v^2}\ |H|^2 G^a_{\mu \nu} G^{a \mu\nu} \ ,
  \eeqn
where $g_s$ is the QCD coupling constant, for which
\beq
       c_H ={3\over (4\pi)^2} {4 v^2 \over 5 M^2} (y_U^2 + y_D^2)^2 \ , \qquad
       c_{GG} = -{1\over (4\pi)^2} {v^2 \over 3M^2}(y_U^2 + y_D^2) \ .
\eeqn
These are small effects if the Yukawa couplings $y_U$ and $y_D$ are
small, but there is no reason for this.   In Little Higgs models, for
example, the top quark Yukawa coupling is given by
\beq
                y_t^2  \approx 1  = { y_1^2 y_2^2 \over  y_1^2 + y_2^2}
                \ , 
\eeqn
forcing both underlying Yukawa couplings to be greater than 1.
Even a value   $y_U = 3$ still corresponds to $\alpha_U = 0.7$ and give
relatively small shifts of TeV-scale vectorlike quark masses.  It is
true that 
the calculation of $c_{GG}$ in Little Higgs models typically leads to
cancellations due to the extra symmetries of the Little Higgs
mechanism.  To see how this works in explicit models, 
see \cite{Han:2003gf,Hubisz:2005tx,Chen:2006cs}.  These models do
still give a search reach beyond the HL-LHC value of 1.5~TeV quoted above.

\begin{figure}
\begin{center}
\includegraphics[width=0.8\hsize]{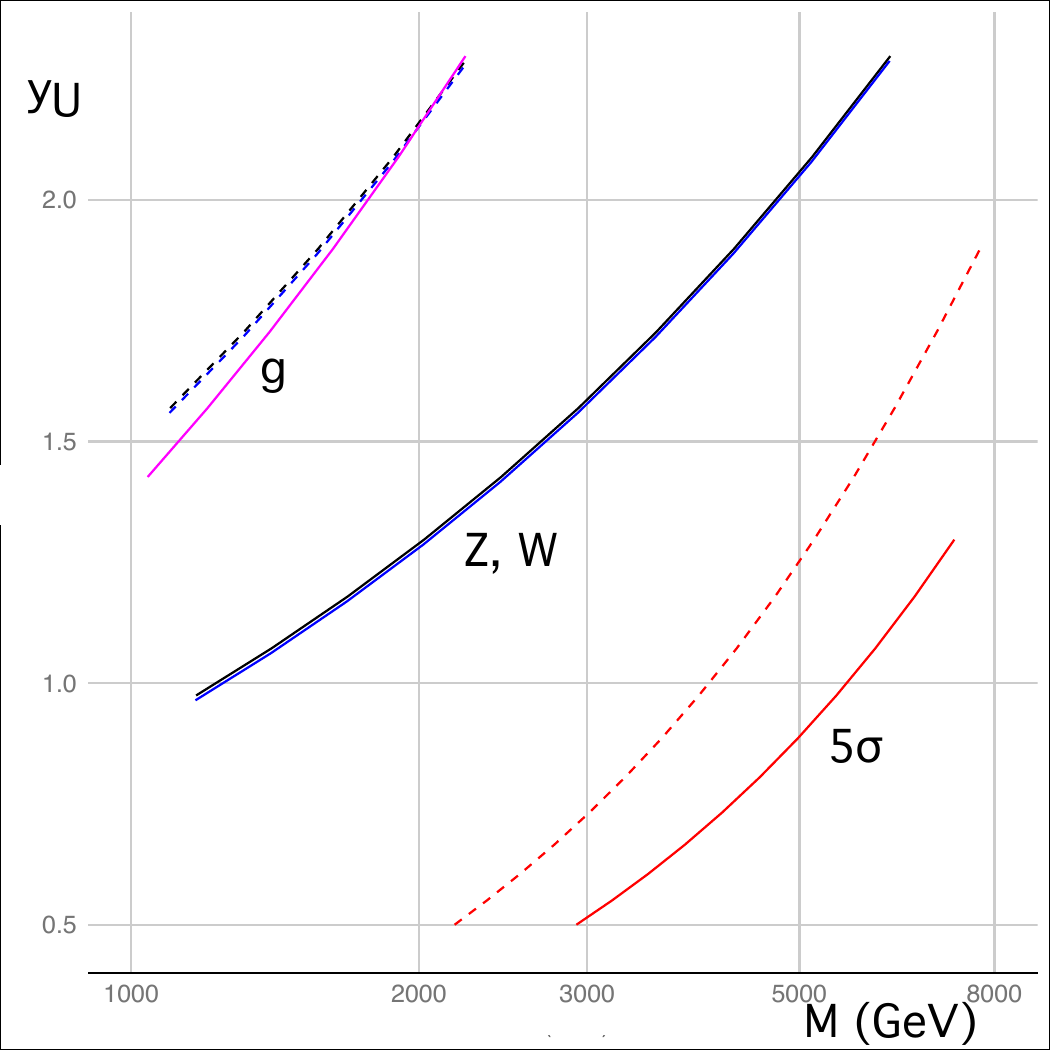}
\end{center}
\caption{Mass reach for the deviation in the Higgs boson couplings due
  to integrating out a vectorlike quark doublet. 
  The blue, black, and magenta curves indicate the 3~$\sigma$
  contours for the deviation in the $HWW$, $HZZ$, and $Hgg$ couplings 
in a SMEFT fit to projected ILC
  data up to 500~GeV.   The curves are plotted in the $(M, y_U)$
  plane for $y_D = 0$ (dashed lines) and $y_D = y_U$ (solid lines). 
  The
  red curves show the 5~$\sigma$ contour for the full SMEFT fit.}
\label{fig:vectorlike}
\end{figure}

    In Fig.~\ref{fig:vectorlike}, I plot the contours in the ($M, y_U$)
plane along which individual Higgs boson couplings have a
  3~$\sigma$ deviation from the Standard Model expectation, assuming
  measurements with the uncertainties expected for the 500~GeV ILC.
   The
  combination of these effects and others gives the contour for a
  5~$\sigma$ deviation from the Standard Model which is also shown in
  the figure.  

 \section{The $t$ squark-Higgsino system}

Finally, as an example of a much more model-specific diagram that can
lead to a significant
effect, I discuss the stop-Higgsino loop correction to the $b$ quark
Yukawa coupling.  This is enhanced by
the top quark Yukawa coupling, the (possibly heavy) higgsino mass, and the
top quark $A$ term in
the numerator.  The correction is generated from the top quark Yukawa
coupling but corrects the much smaller bottom quark Yukawa coupling.
 The effect occurs on top of  other supersymmetry signals, such as
those from the THDM structure, and it  depends on its own specific
parameter set.  Thus, it is very difficult to understand the
importance of this effect from general parameter scans of the MSSM
model space.  Here I give an  estimate of  its potential size, though it should be
emphasized that treating this term in isolation is an 
oversimplification. 

Begin from  the diagram 
\beq
\includegraphics[width=0.4\hsize]{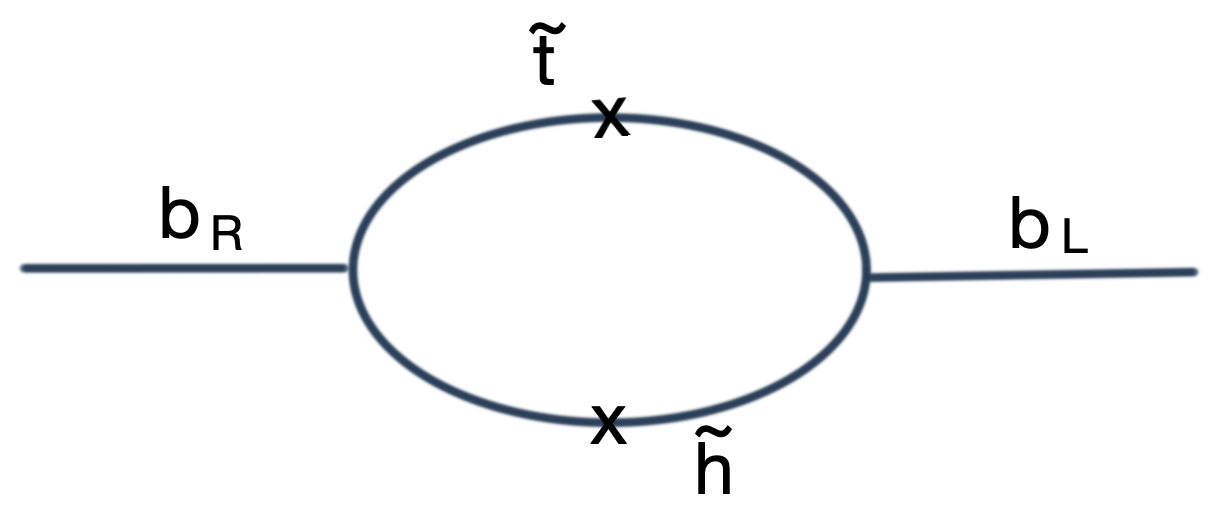}
\eeq{thdiagram}
which gives a loop correction to the $b$ quark  mass or the $b$ quark
Yukawa coupling.  In principle, this diagram and the corresponding
diagram with a $b$ squark could generate a large part of the $b$ quark
mass. Thus, this example illustrates the possibility of large
corrections to the Higgs couplings in models withmass generation by
radiative feed-down.  To give a nonzero effect, the diagram requires a
helicity flip in the numerator of the Higgsino propagator, and a flip
from $\tilde t_R$ to $\tilde t_L$ induced by the top squark $A$ term.
When we expand this diagram in powers of a background field $H$, it
also generates the dimension-6 operator ${\cal O}_b$ that corrects the
$b$ quark Yukawa.  Note that, for  the similar diagram with a $b$
squark,  the dimension-6 contribution is proportional to $m_b^3$
and thus is smaller by $(m_b/m_t)^2$.

\begin{figure}
\begin{center}
\includegraphics[width=0.8\hsize]{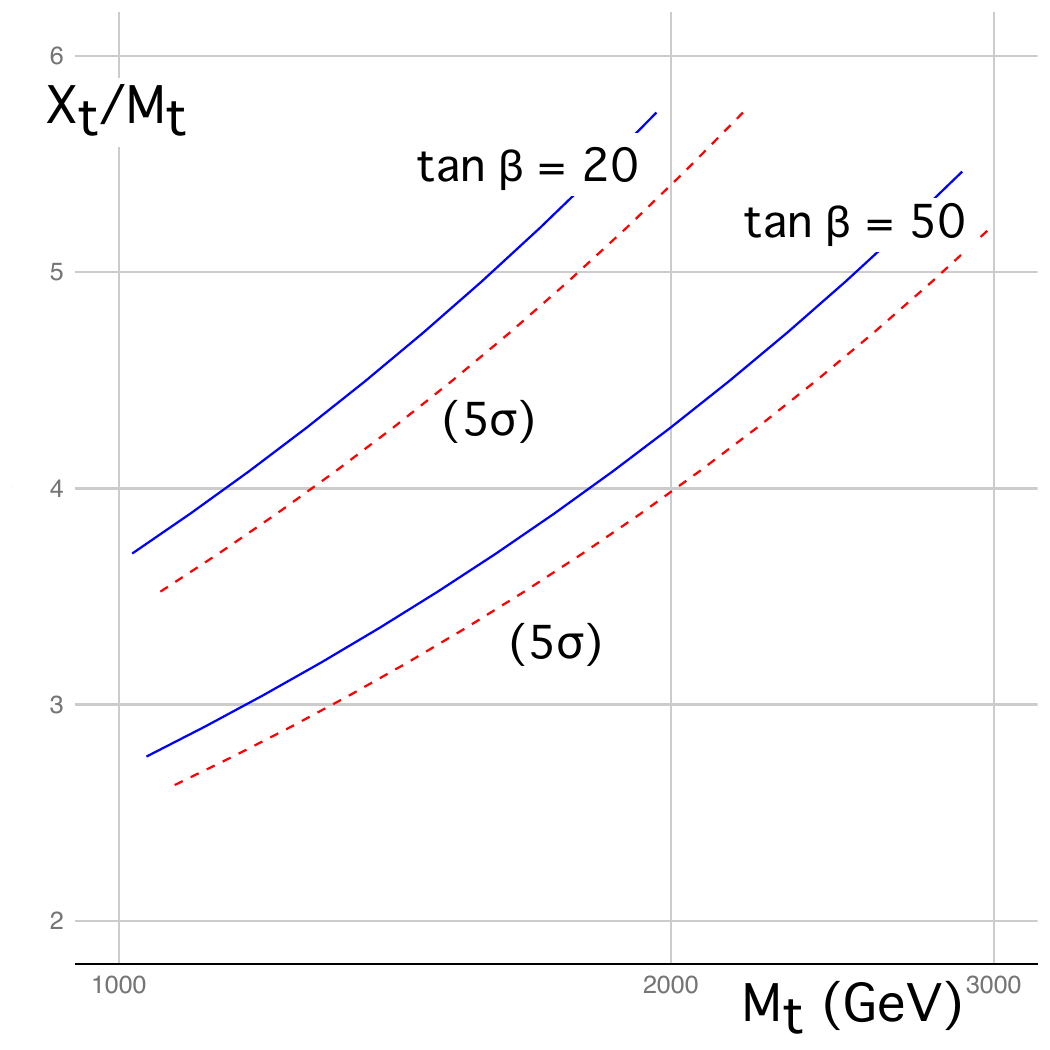}
\end{center}
\caption{Mass reach for the deviation in the Higgs boson coupling to
  $b$  due to the
  diagram \leqn{thdiagram}.   The blue curves indicate the 3~$\sigma$
  contours for the deviation in the $Hbb$ coupling  in a SMEFT fit to projected ILC
  data up to 500~GeV.   The curves are plotted in the $(M, X_t/M)$
  plane for $\tan \beta = 20, 50$.
  The
  red curves show the 5~$\sigma$ contour for the full SMEFT fit.}
\label{fig:thlike}
\end{figure}

In evaluating the diagram in \leqn{thdiagram},  I make some
simplifications.  Since we are interested in heavy masses in the loop,
I ignore terms in the $\tilde t$ and $\tilde h$ (mass$^2$) matrices
proportional to $m_Z^2$ and $m_W^2$. This, in particular, sets the
charged Higgsino mass to $\mu$, which I take to be large.  I also set the soft
supersymmetry breaking masses of the  $\tilde t_R$ and  $\tilde t_L$
to be equal, with the value $M$. Then the top squark mass matrix is 
\beq
      {\cal M}^2 = \pmatrix{ M^2     &    m_t X_t \cr   m_t X_t  &
        M^2 \cr} \ , 
\eeqn
where $X_t = A_t - \mu \cot \beta$. 
I run $y_t(Q)$ to the scale $Q= M$;
the running of $y_b$ from $m_h$ to $M$ cancels that of the generated
dimension-6 operator running from $M$ to $m_h$.   With these  simplifications, the
diagram leads to the effective operator
\beq
\Delta \L = -\Delta m_b \ \bar b_L \cdot H \, b_R 
\eeqn
where 
\beq
\Delta m_b = i { y_b y_t \over \sin\beta\cos\beta} \int {d^4p\over (2\pi)^4} {\mu \ m_t X_t \over 
       (p^2 - \mu^2) ( (p^2 - M^2)^2 - (m_t X_t)^2 ) }
\eeqn
and $X_t = A_t - \mu \cot\beta$. This operator is a function of the
constant background Higgs field $H$.   Expanding in $H$ reveals the
correction to be a sum over operators of dimension 4, 6, etc.
Carrying out this expansion, we find
\beq
\Delta \L  = -  \Delta y_b  \bar b_L \cdot H b_R - {c_b\over v^2 }\,
y_b |H|^2
\bar b_L \cdot H b_R - \cdots \ , 
\eeqn
with 
\beqa
  \Delta y_b &=&{ y_b y^2_t  \mu \over \sin\beta\cos\beta}  X_t\cdot (-i)  \int { d^4p\over (2\pi)^4}{1\over (p^2
    - \mu^2)( p^2 - M^2)^2} \CR
  c_b/v^2  =   &=& { y_b y_t^2 \mu X_t^3 \over \sin\beta\cos\beta} \cdot (-i)  \int { d^4p\over (2\pi)^4}{1\over (p^2
    - \mu^2)( p^2 - M^2)^4} 
\eeqan
In particular,
\beqa
    c_b & =& -  v^2 \cdot  { 3\over 10}  { y_t^2 \mu X_t^3\over (4\pi)^2 \sin\beta\cos\beta}  \CR
          &   & \hskip 0.5in \cdot \biggl[ { \mu^6 - 6 M^2 \mu^4 + 3 M^4
          \mu^2 + 2 M^6 + 6M^4\mu^2 \log ( \mu^2/M^2) \over 2 M^4
          (\mu^2 - M^2)^4 }\biggr]
\eeqa{cbexpress}
This expression predicts the largest effects when $\mu^2 > M^2$, with
both being much greater than $v^2$.

    In Fig.~\ref{fig:thlike}, I plot the contours in the ($M, X_t$)
plane along which individual Higgs boson couplings have a
  3~$\sigma$ deviation from the Standard Model expectation, assuming
  measurements with the uncertainties expected for the 500~GeV ILC.
  Other Higgs factory proposals give very similar expectations.  The
  contours are given for $M/\mu = 0.5$ and $\tan \beta = 20$, 50.
 The needed values of $X_t$ are large but not excessively so.
 According to \cite{Kusenko:1996jn}, the instability to a color and
 charge breaking minimum is unimportant for such large values of the
 superpartner
masses.    The deviations due
  to this effect would typically add to deviations induced by other 
supersymmetry effects.  For this reason, I draw the 5 $\sigma$
contours with dashed lines.

 \section{Conclusions}  

 In this paper, I have shown that
  models of new physics can have signficant effects on the Higgs boson
  couplings even when the masses of the new particles that induce
  these effects are well beyond the direct search reach of the
  HL-LHC.   This does not happen in all models, but it opens new
  classes of models and new regions of parameter space to exploration.

It is noteworthy that each model leads to its own pattern of
deviations in the Higgs boson couplings, as was stressed in
\cite{Barklow:2017suo}.
This implies that a discovery of deviations will also give information
about the nature of the new physics that causes it.

 In the best case, the regions of parameter space made accessible by
 precision Higgs measurements are beyond the scope of LHC searches
 today but within the scope of searches at the HL-LHC.   Then
 discoveries at the HL-LHC and the Higgs factory can be truly
 complementary, reinforcing one another and offering different
 viewpoints on the nature of the new physics. 

\Acknowledgements

The results presented in this paper are part of a more general
analysis in progress in collaboration with Ian Banta and Nathaniel
Craig.  I thank Ian and Nathaniel for suggesting this approach to the
understanding of precision Higgs boson coupling mass reach, and for 
discussion of the results presented here.  This analysis also
benefited from insights and criticism from 
Christophe Grojean, Mihoko Nojiri,  and Maxim
Perelstein and from Sally Dawson,  Patrick Meade, Laura Reina, and Caterina
Vernieri.   My work is supported by the US Department of Energy under
                     contract DE--AC02--76SF00515.


\begin{thebibliography}{99}

  
\bibitem{Barklow:2017suo}
T.~Barklow, K.~Fujii, S.~Jung, R.~Karl, J.~List, T.~Ogawa, M.~E.~Peskin and J.~Tian,
``Improved Formalism for Precision Higgs Coupling Fits,''
Phys. Rev. D \textbf{97}, 053003 (2018)
[arXiv:1708.08912 [hep-ph]].

\bibitem{ILCInternationalDevelopmentTeam:2022izu}
A.~Aryshev \textit{et al.} [ILC International Development Team],
``The International Linear Collider: Report to Snowmass 2021,''
[arXiv:2203.07622 [physics.acc-ph]].


\bibitem{Cahill-Rowley:2014wba}
M.~Cahill-Rowley, J.~Hewett, A.~Ismail and T.~Rizzo,
``Higgs boson coupling measurements and direct searches as complementary probes of the phenomenological MSSM,''
Phys. Rev. D \textbf{90}, 095017 (2014)
[arXiv:1407.7021 [hep-ph]].

\bibitem{Djouadi:1997yw}
A.~Djouadi, J.~Kalinowski and M.~Spira,
``HDECAY: A Program for Higgs boson decays in the standard model and its supersymmetric extension,''
Comput. Phys. Commun. \textbf{108}. 56  (1998).
[arXiv:hep-ph/9704448 [hep-ph]],
 \url{http://tiger.web.psi.ch/hdecay/}.

\bibitem{CidVidal:2018eel}
X.~Cid Vidal, M.~D'Onofrio, P.~J.~Fox, R.~Torre, K.~A.~Ulmer, A.~Aboubrahim, A.~Albert, J.~Alimena, B.~C.~Allanach and C.~Alpigiani, \textit{et al.}
``Report from Working Group 3: Beyond the Standard Model physics at the HL-LHC and HE-LHC,''
CERN Yellow Rep. Monogr. \textbf{7}, 585  (2019)
[arXiv:1812.07831 [hep-ph]].



\bibitem{LCCPhysicsWorkingGroup:2019fvj}
K.~Fujii \textit{et al.} [LCC Physics Working Group],
[arXiv:1908.11299 [hep-ex]].

\bibitem{Barklow:2017awn}
T.~Barklow, K.~Fujii, S.~Jung, M.~E.~Peskin and J.~Tian,
``Model-Independent Determination of the Triple Higgs Coupling at e+e- Colliders,''
Phys. Rev. D \textbf{97}, 053004 (2018)
[arXiv:1708.09079 [hep-ph]].



\bibitem{Giudice:2007fh}
G.~F.~Giudice, C.~Grojean, A.~Pomarol and R.~Rattazzi,
``The Strongly-Interacting Light Higgs,''
JHEP \textbf{06}, 045 (2007)
[arXiv:hep-ph/0703164 [hep-ph]].

\bibitem{EuropeanStrategyforParticlePhysicsPreparatoryGroup:2019qin}
R.~K.~Ellis, B.~Heinemann, J.~de Blas, M.~Cepeda, C.~Grojean, F.~Maltoni, A.~Nisati, E.~Petit, R.~Rattazzi and W.~Verkerke, \textit{et al.}
``Physics Briefing Book: Input for the European Strategy for Particle Physics Update 2020,''
[arXiv:1910.11775 [hep-ex]].

\bibitem{Bernon:2015qea}
J.~Bernon, J.~F.~Gunion, H.~E.~Haber, Y.~Jiang and S.~Kraml,
``Scrutinizing the alignment limit in two-Higgs-doublet models: m$_h$=125  GeV,''
Phys. Rev. D \textbf{92},   075004 (2015) 
[arXiv:1507.00933 [hep-ph]].

\bibitem{Haber:1994mt}
H.~E.~Haber,
``Nonminimal Higgs sectors: The Decoupling limit and its phenomenological implications,''
[arXiv:hep-ph/9501320 [hep-ph]].

\bibitem{Egana-Ugrinovic:2015vgy}
D.~Egana-Ugrinovic and S.~Thomas,
``Effective Theory of Higgs Sector Vacuum States,''
[arXiv:1512.00144 [hep-ph]].

\bibitem{Belusca-Maito:2016dqe}
H.~B\'elusca-Ma\"\i{}to, A.~Falkowski, D.~Fontes, J.~C.~Rom\~ao and J.~P.~Silva,
``Higgs EFT for 2HDM and beyond,''
Eur. Phys. J. C \textbf{77}, no.3, 176 (2017)
[arXiv:1611.01112 [hep-ph]].


\bibitem{CMS:2019qzn}
 CMS Collaboration, 
``Search for a new scalar resonance decaying to a pair of Z bosons at the High-Luminosity LHC,''
CMS-PAS-FTR-18-040 (2019).




\bibitem{Banta:2021dek}
I.~Banta, T.~Cohen, N.~Craig, X.~Lu and D.~Sutherland,
``Non-decoupling new particles,''
JHEP \textbf{02}, 029 (2022)
[arXiv:2110.02967 [hep-ph]].

\bibitem{Henning:2014wua}
B.~Henning, X.~Lu and H.~Murayama,
``How to use the Standard Model effective field theory,''
JHEP \textbf{01}, 023 (2016)
[arXiv:1412.1837 [hep-ph]].

\bibitem{CLIC:2018fvx}
J.~de Blas \textit{et al.} [CLIC],
``The CLIC Potential for New Physics,''
[arXiv:1812.02093 [hep-ph]].


\bibitem{Huang:2016cjm}
P.~Huang, A.~J.~Long and L.~T.~Wang,
``Probing the Electroweak Phase Transition with Higgs Factories and Gravitational Waves,''
Phys. Rev. D \textbf{94}, no.7, 075008 (2016)
[arXiv:1608.06619 [hep-ph]].



\bibitem{DiVita:2017eyz}
S.~Di Vita, C.~Grojean, G.~Panico, M.~Riembau and T.~Vantalon,
``A global view on the Higgs self-coupling,''
JHEP \textbf{09}, 069 (2017)
[arXiv:1704.01953 [hep-ph]].

\bibitem{Durieux:2022hbu}
G.~Durieux, M.~McCullough and E.~Salvioni,
``Charting the Higgs self-coupling boundaries,''
[arXiv:2209.00666 [hep-ph]].


\bibitem{Angelescu:2020yzf}
A.~Angelescu and P.~Huang,
``Integrating Out New Fermions at One Loop,''
JHEP \textbf{01}, 049 (2021)
[arXiv:2006.16532 [hep-ph]].



\bibitem{Han:2003gf}
T.~Han, H.~E.~Logan, B.~McElrath and L.~T.~Wang,
``Loop induced decays of the little Higgs: H ---\ensuremath{>} gg, gamma gamma,''
Phys. Lett. B \textbf{563}, 191 (2003)
[erratum: Phys. Lett. B \textbf{603}, 257 (2004)]
[arXiv:hep-ph/0302188 [hep-ph]].

\bibitem{Hubisz:2005tx}
J.~Hubisz, P.~Meade, A.~Noble and M.~Perelstein,
``Electroweak precision constraints on the littlest Higgs model with T parity,''
JHEP \textbf{01}, 135 (2006)
[arXiv:hep-ph/0506042 [hep-ph]].

\bibitem{Chen:2006cs}
C.~R.~Chen, K.~Tobe and C.~P.~Yuan,
``Higgs boson production and decay in little Higgs models with T-parity,''
Phys. Lett. B \textbf{640}, 263 (2006)
[arXiv:hep-ph/0602211 [hep-ph]].

\bibitem{Kusenko:1996jn}
A.~Kusenko, P.~Langacker and G.~Segre,
``Phase transitions and vacuum tunneling into charge and color breaking minima in the MSSM,''
Phys. Rev. D \textbf{54}, 5824  (1996)
[arXiv:hep-ph/9602414 [hep-ph]].



\end{thebibliography}
\end{document}